\documentstyle[preprint,aps,epsf]{revtex}
\tighten
\draft

\begin{document}
\title{The effect of surface and Coulomb interaction on the liquid-gas phase
transition of finite nuclei}
\author{W.L.Qian$^{a,b}$ Ru-Keng Su$^{c,a}$ H.Q.Song$^{c,d}$}
\address{$^a$Department of Physics, Fudan University, Shanghai 200433,\\
P.R.China}
\address{$^b$Surface Physics Lab. (National Key Lab.) Fudan University\\
Shanghai 200433, P.R.China}
\address{$^c$China Center of Advanced Science and Technology\\
(World Laboratory)\\
P.O.Box 8730, Beijing 100080, P.R.China}
\address{$^d$Shanghai Institute of Nuclear Research, China Academy of Science\\
P.O.Box 800204, Shanghai 201800, P.R.China}
\maketitle

\begin{abstract}
By means of the Furnstahl, Serot and Tang's model, the effects of surface
tension and Coulomb interaction on the liquid-gas phase transition for
finite nuclei are investigated. A limit pressure $p_{\lim }$ above which the
liquid-gas phase transition cannot take place has been found. It is found
that comparing to the Coulomb interaction, the contribution of surface
tension is dominate in low temperature regions. The binodal surface is also
addressed.
\end{abstract}


\newpage

Since the arguments given by M\"{u}ller and Serot [1] that a second order
liquid-gas (L-G) phase transition will take place in a multi-components and
multi-conserved charged system, much theoretical efford has been devoted for
studying this problem by using different models and different treatments
[2-4]. But all investigations are limited to infinite nuclear matter. It is
of interest to extend this study to finite nuclei. This is the objective of
this paper.

If we consider the finite nuclei as a liquid droplet and discuss its L-G
phase transition, two major effects, namely, the surface energy of the
droplet and the Coulomb interaction of proton-proton must be considered. The
reasons are as follows: it has been shown that the difference of the
chemical potentials between proton and neutron plays an essential role to
determine the order of L-G phase transition [1, 2]. For the infinite nuclear
matter the chemical potentials of proton and neutron depend on the third
component $I_3$ of isospin when nucleon-nucleon-$\rho $-meson ($NN\rho $)
interaction exists. In symmetric nuclear matter, the L-G phase transition is
of first order because $I_3=0$. In asymmetric nuclear matter, $I_3\neq 0$
and then the chemical potential of neutron $\mu _n$ does not equal to the
chemical potential of proton $\mu _p$, a second order phase transition may
take place. The Coulomb interaction cannot be taken into account because it
becomes divergent in {\it infinite} nuclear matter. But in {\it finite}
nuclei, the contribution of Coulomb interaction can be considered.
Obviously, the chemical potential of proton $\mu _p$ not only depends on $I_3
$, but also on Coulomb interaction. But the later has no effect on $\mu _n$.
The contribution of Coulomb interaction will make that the values of $\mu _p$
and $\mu _n$ becomes more different.

Besides Coulomb interaction, on the other hand, the surface tension of the
droplet will affect on the pressure of the liquid phase and then on the
coexistence equations because the pressures of two phases must equal at the
phase transition point.

To exhibit the effects of surface energy and Coulomb interaction on the L-G
phase transition for finite nuclei, we employ the Furnstahl-Serot-Tang (FST)
model [5-8], which has been shown to be successful to explain the properties
of both infinite nuclear matter and finite nuclei. The Lagrangian density of
FST model under mean field approximations reads 
\begin{eqnarray}
L_{MFT} &=&\overline{\Psi }\left[ i\gamma ^\mu \partial _\mu -\left(
M-g_s\phi _0\right) -g_v\gamma ^0V_0-\frac 12g_\rho \tau _3\gamma
^0b_0\right] \Psi  \label{e1} \\
&&+\frac 12m_v^2V_0^2\left( 1+\eta \frac{\phi _0}{S_0}\right) +\frac 1{4!}
\zeta \left( g_vV_0\right) ^4+\frac 12m_\rho ^2b_0^2  \nonumber \\
&&-H_q\left( 1-\frac{\phi _0}{S_0}\right) ^{4/d}\left[ \frac 1dln\left( 1- 
\frac{\phi _0}{S_0}\right) -\frac 14\right]  \nonumber
\end{eqnarray}
where $g_s$, $g_v$ $g_\rho $ are, respectively, the couplings of light
scalar meson $\sigma $, vector meson $\omega $ and isovector meson $\rho $
fields to the nucleon, $\phi _0$, V$_0$, b$_0$ are the expectation values $
\phi _0\equiv <\phi >$, $<V_\mu >\equiv \delta _{\mu 0}V_0$, $<b_{\mu
3}>\equiv \delta _{\mu 0}b_0$. The scalar fluctuation field $\phi $ is
related to $S$ by $S\left( x\right) =S_0-\phi \left( x\right) $ and $H_q$ is
given by $m_s^2=4H_q/\left( d^2S_0^2\right) $, d the scalar dimension. By
using the standard technique of statistical mechanics, we get the
thermodynamic potential $\Omega $ as [2, 9] 
\begin{eqnarray}
\Omega &=&V\left\{ H_q\left[ \left( 1-\frac{\phi _0}{S_0}\right)
^{4/d}\left( \frac 1dln\left( 1-\frac{\phi _0}{S_0}\right) -\frac 14\right)
+ \frac 14\right] \right.  \label{e2} \\
&&\left. -\frac 12m_\rho ^2b_0^2-\frac 12\left( 1+\eta \frac{\phi _0}{S_0}
\right) m_v^2V_0^2-\frac 1{4!}\zeta \left( g_vV_0\right) ^4\right\} 
\nonumber \\
&&-2k_BT\left[ \sum_{k,\tau }ln\left( 1+e^{-\beta \left( E^{*}\left(
k\right) -\nu _\tau \right) }\right) +\sum_{k,\tau }ln\left( 1+e^{-\beta
\left( E^{*}\left( k\right) +\nu _\tau \right) }\right) \right]  \nonumber
\end{eqnarray}
where $\beta =1/k_BT$ and the quantity $\nu _i$ $\left( i=n,p\right) $ is
related to the usual chemical potential $\mu _i$ by the equations 
\begin{equation}
\nu _n=\mu _n-g_vV_0+\frac{g_\rho ^2\rho _3}{4m_\rho ^2}  \label{e3}
\end{equation}
\begin{equation}
\nu _p=\mu _p-g_vV_0-\frac{g_\rho ^2\rho _3}{4m_\rho ^2}  \label{e4}
\end{equation}
where $\rho _3=\rho _p-\rho _n$, the third component of isospin $I_3=\left(
N_p-N_n\right) /2=V\rho _3/2$, and $E^{*}\left( k\right) =\sqrt{M^{*2}+k^2}$
with $M^{*}=M-g_s\phi _0$. Usually, instead of $\rho _3$, we introduce $
\alpha =\left( \rho _n-\rho _p\right) /\rho $, the asymmetric parameter to
calculate, where $\rho =\rho _n+\rho _p$. Having obtained the thermodynamic
potential, all other thermodynamic quantities, for example, pressure $p$,
can be calculated. We get 
\begin{eqnarray}
p &=&-H_q\left\{ \left( 1-\frac{\phi _0}{S_0}\right) ^{\frac 4d}\left[ \frac 
1dln\left( 1-\frac{\phi _0}{S_0}\right) -\frac 14\right] +\frac 14\right\}
\label{e5} \\
&&+\frac 12\left( 1+\eta \frac{\phi _0}{S_0}\right) m_v^2V_0^2+\frac 1{4!}
\zeta g_v^4V_0^4+\frac{g_\rho ^2}{8m_\rho ^2}\rho _3^2  \nonumber \\
&&+\frac 13\frac 2{(2\pi )^3}\int d^3k\frac{k^2}{E^{*}(k)}[n_n(k)+\overline{
n }_n(k)+n_p(k)+\overline{n}_p(k)].  \nonumber
\end{eqnarray}
where 
\begin{eqnarray}
n_\tau (k) &=&\{exp[(E^{*}(k)-\nu _\tau )/k_BT]+1\}^{-1}  \label{e6} \\
\overline{n}_\tau (k) &=&\{exp[(E^{*}(k)+\nu _\tau )/k_BT]+1\}^{-1}
\label{e7}
\end{eqnarray}
\[
(\tau =n,p) 
\]
are the nucleon and anti-nucleon distributions respectively. The neutron
density $\rho _n$ and the proton density $\rho _p$ are given by 
\begin{equation}
\rho _\tau =\frac 2{(2\pi )^3}\int d^3k[n_\tau (k)-\overline{n}_\tau (k)]. 
\hspace{2cm}(\tau =n,p)  \label{e8}
\end{equation}

Now we are in a position to study the L-G phase transition of finite nuclei.
The two-phase coexistence equations are 
\begin{eqnarray}
\mu _n^{\prime }\left( T,\rho ^{\prime },\alpha ^{\prime }\right) &=&\mu
_n^{\prime \prime }\left( T,\rho ^{\prime \prime },\alpha ^{\prime \prime
}\right)  \label{e9} \\
\mu _p^{\prime }\left( T,\rho ^{\prime },\alpha ^{\prime }\right) +\mu
_{Coul}\left( \rho ^{\prime }\right) &=&\mu _p^{\prime \prime }\left( T,\rho
^{\prime \prime },\alpha ^{\prime \prime }\right)  \label{e10}
\end{eqnarray}
\begin{equation}
p^{\prime }\left( T,\rho ^{\prime },\alpha ^{\prime }\right)
+p_{Coul}^{\prime }\left( \rho ^{\prime }\right) +p_{surf}^{\prime }\left(
T,\rho ^{\prime }\right) =p^{\prime \prime }\left( T,\rho ^{\prime \prime
},\alpha ^{\prime \prime }\right)  \label{e11}
\end{equation}
where the prime and the double prime refer to the liquid phase and gas
phase, respectively. Considering the droplet as an uniformly charged sphere,
the contribution of Coulomb interaction to the chemical potential of proton
and the pressure are [9, 10, 11]. 
\begin{equation}
\mu _{Coul}=\frac 65\frac{Ze^2}R  \label{e12}
\end{equation}
\begin{equation}
p_{Coul}\left( \rho \right) =\frac{Z^2e^2}{5AR}\rho  \label{e13}
\end{equation}
respectively. The additional pressure provided by the surface tension of the
liquid droplet is [12, 13] 
\begin{equation}
p_{surf}\left( T,\rho \right) =-2\gamma \left( T\right) /R  \label{e14}
\end{equation}
where 
\begin{equation}
\gamma \left( T\right) =\left( 1.14{\bf MeV}fm^{-2}\right) \left[ 1+\frac{3T 
}{2T_c}\right] \left[ 1-\frac T{T_c}\right] ^{3/2}  \label{e15}
\end{equation}
with $T_c$ being the critical temperature of L-G phase transition in
symmetric nuclear matter. We take the liquid droplet along the $\beta $
-stability line, it satisfies 
\begin{equation}
Z=0.5A-0.3\times 10^{-2}A^{5/3}  \label{e16}
\end{equation}
The parameters be chosen for our numerical calculations are the set $T_1$ of
FST model 
\begin{eqnarray}
g_s^2 &=&99.3,\hspace{5mm}g_v^2=154.5,\hspace{5mm}g_\rho ^2=70.2  \label{e17}
\\
m_s &=&509MeV{\bf ,}\hspace{5mm}S_0=90.6MeV  \nonumber \\
\zeta &=&0.0402,\hspace{5mm}\eta =-0.496,\hspace{5mm}d=2.70  \nonumber
\end{eqnarray}

Our results are summerized in Fig.1-Fig.6. To make our results more
transparent, we neglect the Coulomb interaction and the surface effect by
taking $R\rightarrow \infty $ in Fig.1 and the $\mu _n$, $\mu _p$ isobar vs $
\alpha $ reduce to that of infinite asymmetric matter. In this Figure, we
fix the temperature $T=10$ ${\bf MeV}$ and the curves a, b, c, d, e
correspond to the pressure 0.06, 0.085, 0.100, 0.164 and 0.200 ${\bf MeV}
fm^{-3}$ respectively. We see that the curves for lower pressures are more
complicate than those of the large pressures. When $p=0.200$ ${\bf MeV}
fm^{-3}$, curve e has one branch only, but when $p=0.06$ ${\bf MeV}fm^{-3}$,
curve a has three branchs. The chemical potentials isobar $\mu _n^{\prime
\prime }$, $\mu _p^{\prime \prime }$ vs $\alpha ^{\prime \prime }$ given by
the right hand side of eq.(9-11) for the gas phase is shown in Fig.2 where $
T=5$ ${\bf MeV}$ and $p=0.016$ ${\bf MeV}fm^{-3}$. In fact, these curves are
the same as that of infinite nuclear matter because the chemical potential
and the pressure for the gas phase do not depend on the Coulomb interaction
and the surface term. We see in this case both $\mu _p^{\prime \prime
}\left( \alpha ^{\prime \prime }\right) $ and $\mu _n^{\prime \prime }\left(
\alpha ^{\prime \prime }\right) $ curves have three branchs. The chemical
isobar as a function of $\alpha $ for the liquid phase and the gas phase are
shown in Fig.3 by solid line and dashed line respectively where we fixed $T=5
$ ${\bf MeV}$ and $p=0.016$ ${\bf MeV}fm^{-3}$. The dashed lines in Fig.3
for gas phase are in fact the same curves as those of Fig.2 except that the
range of the $\alpha $-axis is (0.0, 0.5) instead of (0.0, 1.0). The
rectangle construction[1, 2] which represents for the Gibbs' conditions of
eq.(9 - 11) for the two-phase equilibrium is also plotted in Fig.3 by the
edges of a rectangle. Due to the effect of Coulomb interaction and the
surface energy, the chemical potential isobars for the gas phase and for the
liquid phase are very different. We see from Fig.3 that the $\mu _p^{\prime
\prime }\left( \alpha ^{\prime \prime }\right) $ and $\mu _n^{\prime \prime
}\left( \alpha ^{\prime \prime }\right) $ curves for the gas phase have two
branchs shown by dashed lines in the regions $0<\alpha <0.5$ respectively,
but $\mu _p^{\prime }\left( a^{\prime }\right) $ $\mu _n^{\prime }\left(
\alpha ^{\prime }\right) $ for the liquid phase has one branch in this
region only. This behavior is quite different from that of infinite nuclear
matter in which the liquid phase and the gas phase chemical isobars $\mu _n$
and $\mu _p$ are shown by the same curves (for example , see curve a of
Fig.1). This new feature leads to multi-solutions for Gibbs' condition
because one can find four rectangles between two branchs of $\mu _n^{\prime
\prime }\left( \mu _p^{\prime \prime }\right) $ and one branch of $\mu
_n^{\prime }\left( \mu _p^{\prime }\right) $. But according to the
equilibrium condition: the chemical potential of the system in equilibrium
state must take the minimal value at fixed temperature and pressure [14].
Therefore, only one rectangle which corresponds to the minimum chemical
potential isobar shown by solid lines in Fig.3 refers to a stable
equilibrium phase transition, and the others are all metastable states. The
other three rectangles are not shown in Fig.3.

The section of binodal surface [1,2] at finite temperature $T=5$ ${\bf MeV}$
is shown in Fig.4. A limit pressure $p_{\lim }=0.018{\bf MeV}fm^{-3}$ above
which the rectangle cannot be found and the coexistenced equations eq.(9-11)
have no solution has been obtianed. The binodal surface will cut off at
limit pressure $p_{\lim }$. This situation is very similar to that of our
previous paper [2] in which we considered the density dependence of the $
NN\rho $ coupling $g_\rho \left( \rho \right) $. The reason is that no
matter $g_\rho \left( \rho \right) $ or the Coulomb interaction or the
surface energy, even though they change the chemical potential $\mu _n$ and $
\mu _p$ in different fashions, they will make that the rectangle
construction turns out to be disappear. In fact, this result is a reflaction
of the so called Coulomb instability in finite nuclei [9-11, 13, 15]. The
Coulomb instability of FST model has been discussed in detail by our
previous paper[9] in which we found a limit temperature $T_{\lim }$ above
which the coexistenced equations have no solution and the L-G phase
transition can not take place. The difference is that, instead of $T_{\lim }$
, we now fix temperature $T=5$ ${\bf MeV}$ to find the limit pressure.

Finally, we hope to compare the effects of Coulomb interaction and the
surface tension on the L-G phase transition, separably. As shown in eq.(13)
and eq.(14), we see that, firstly, the effects of Coulomb interaction and
the surface tension are opposite because of $p_{Coul}$ and $p_{surf}$ with
opppsite signs, and secondly, $p_{surf}$ depends on temperature but $p_{Coul}
$ is independent. It means that the Coulomb interaction and the surface
tension play different roles in different temperature regions. In low
temperature regions $T{\le }5$ ${\bf MeV}{\ll }T_c$, $\gamma \left( T\right) 
$ becomes larger, and we have $|p_{surf}|>p_{Coul}$, the surface tension
becomes dominant. To show this result clearly, we draw the chemical isobar
vs. $\alpha $ curves with the surface effect only and with the Coulomb
interaction only in Fig.5 and Fig.6 respectively. Comparing Fig.5, Fig.6 and
Fig.3 we see that the curves of Fig.5 are very similar to those of Fig.3,
but the curves of Fig.6 are very different. It confirms the surface effect
dominates at $T=5{\bf MeV}$. The curves in Fig.6 are very complicate. The
reason is that we see from eq.(11), in the phase transition process, $
p_{surf}^{\prime }$ will increase $p^{\prime \prime }$ since it has negative
sign, but $p_{Coul}^{\prime }$ will decrease $p^{\prime \prime }$. As
indicated by Fig.1, the pressure $p^{\prime \prime }$ becomes lower, the
curve becomes more complicate.

In summary, it is shown that the surface effect and the Coulomb interaction
are important for the L-G phase transition of finite nuclei. In low
temperature $T\ll T_c$ regions, the surface effect is dominate. A limit
pressure $p_{\lim }$ above which the L-G phase transition cannot take place
has been found. Since the critical temperature $T_c=14.75$ ${\bf MeV}$, and
the limit temperature $T_{\lim }$ is around $5.4-8.8$ ${\bf MeV}$ for FST
model [12], we come to a conclusion that the surface effect is dominate in
the L-G phase transition of finite nuclei for FST model.

\subsubsection{Acknowledgements}

This work was supported in part by NNSF of China under contract No.
19975010, 19947001, 10075071, the Foundation of Education Ministry of China,
and the Major State Basic Research Development Program in China under
constract No. G200077400.

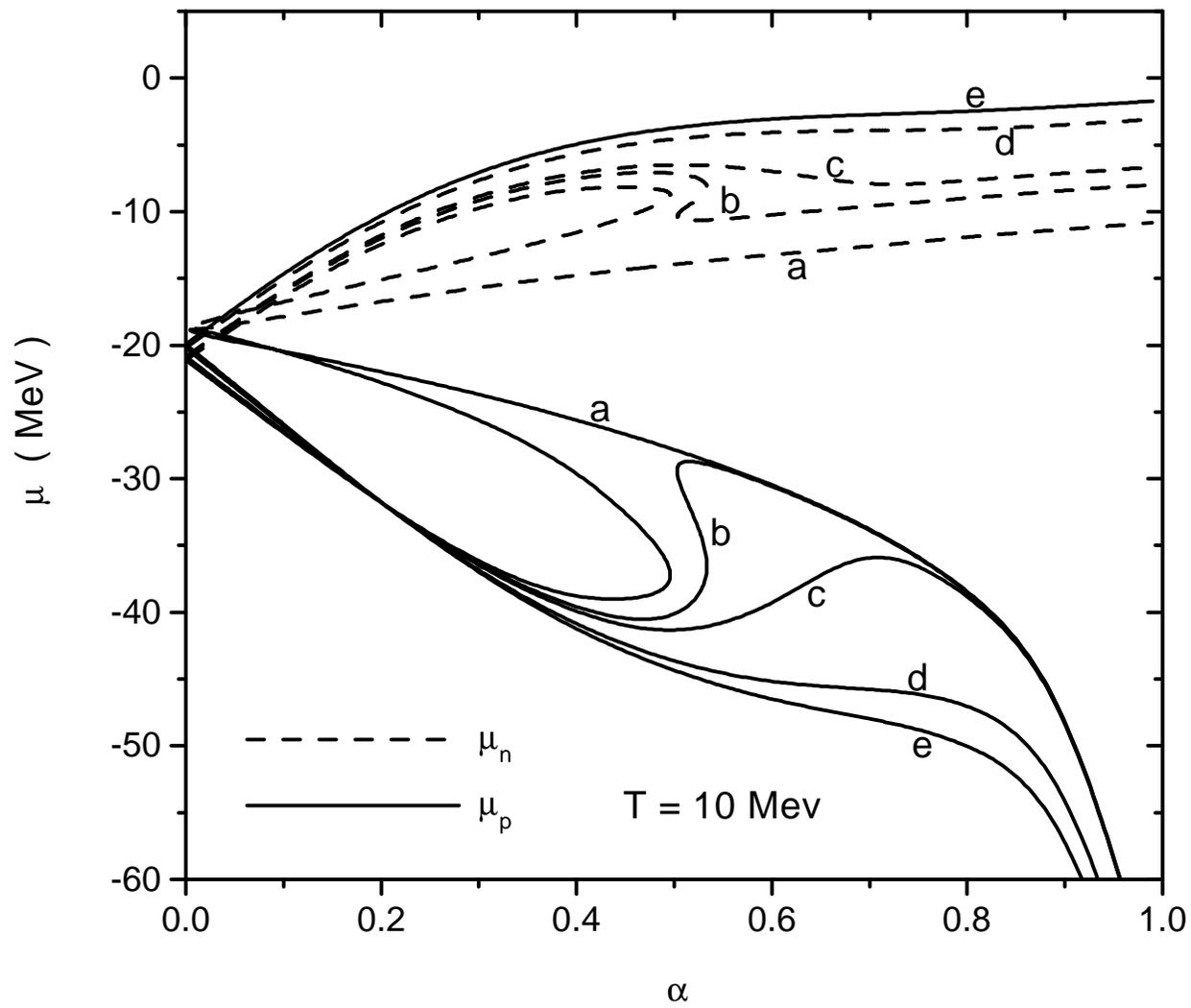
\begin{figure}[h]
\caption{ The chemical isobars for infinite nuclear matter, where $T=10$ $
{\bf MeV}$, and a, b, c, d, e refer to the pressure 0.06, 0.085, 0.100,
0.164 and 0.200${\bf MeV} fm^{-3}$ respectively. }
\label{Fig1}
\end{figure}

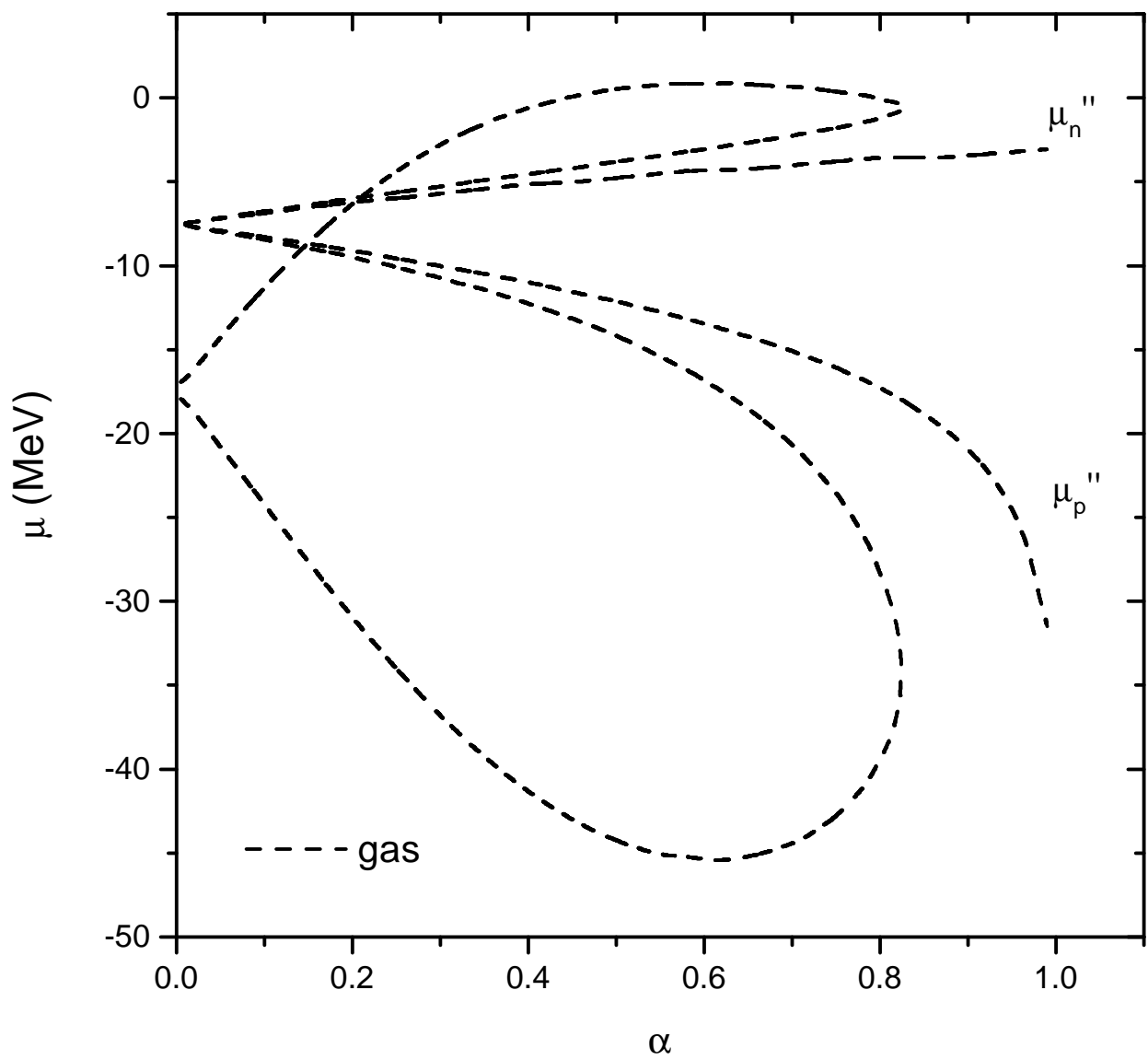
\begin{figure}[h]
\caption{The chemical isobars for the gas phase where $T=5$ ${\bf MeV}$ and $
p=0.016$ ${\bf MeV} fm^{-3}$.}
\label{Fig2}
\end{figure}

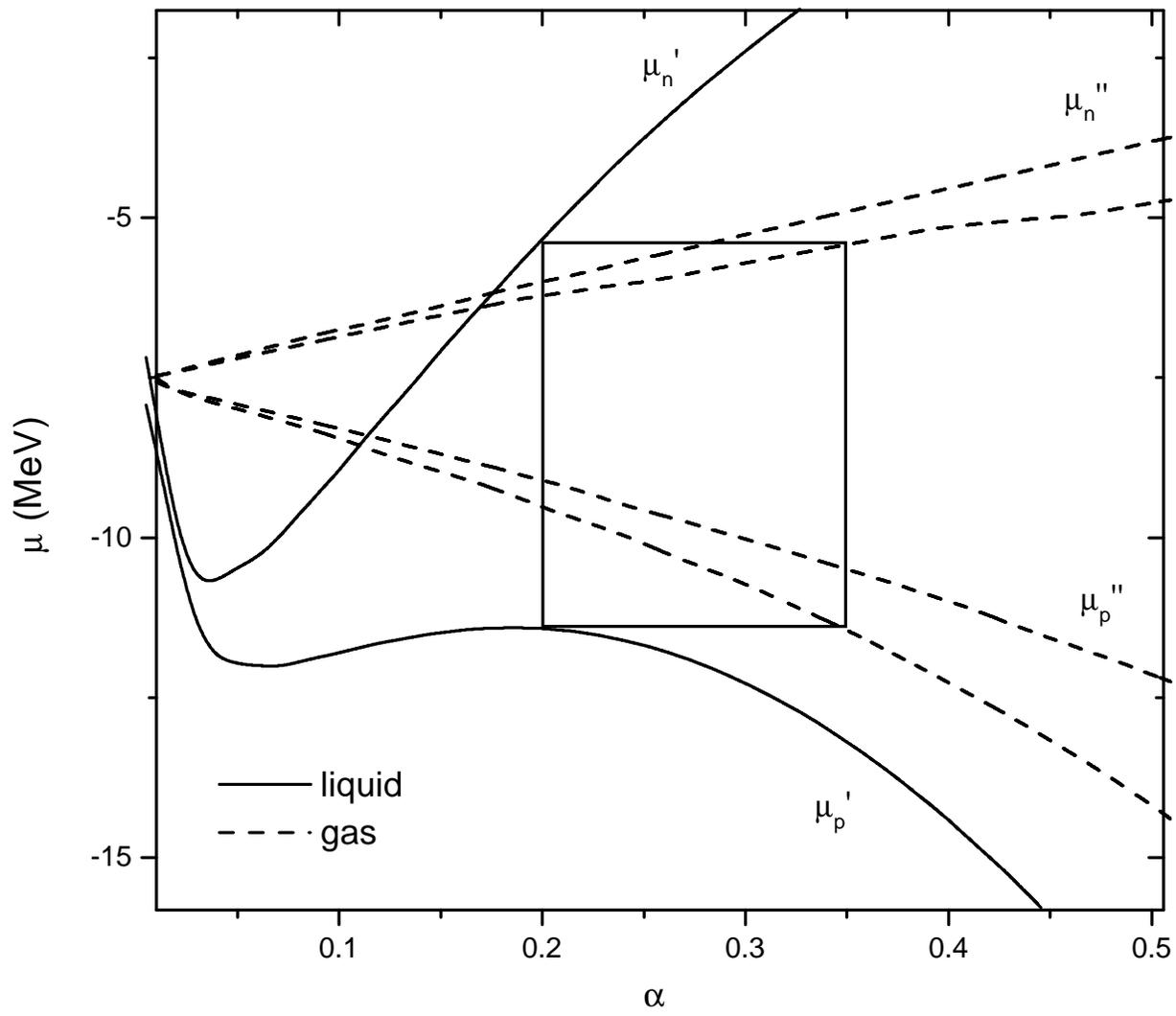
\begin{figure}[h]
\caption{The rectangle construction for two-phase equilibrium for $T=5$ $
{\bf MeV}$ and $p=0.016$ ${\bf MeV}fm^{-3}$.}
\label{Fig3}
\end{figure}

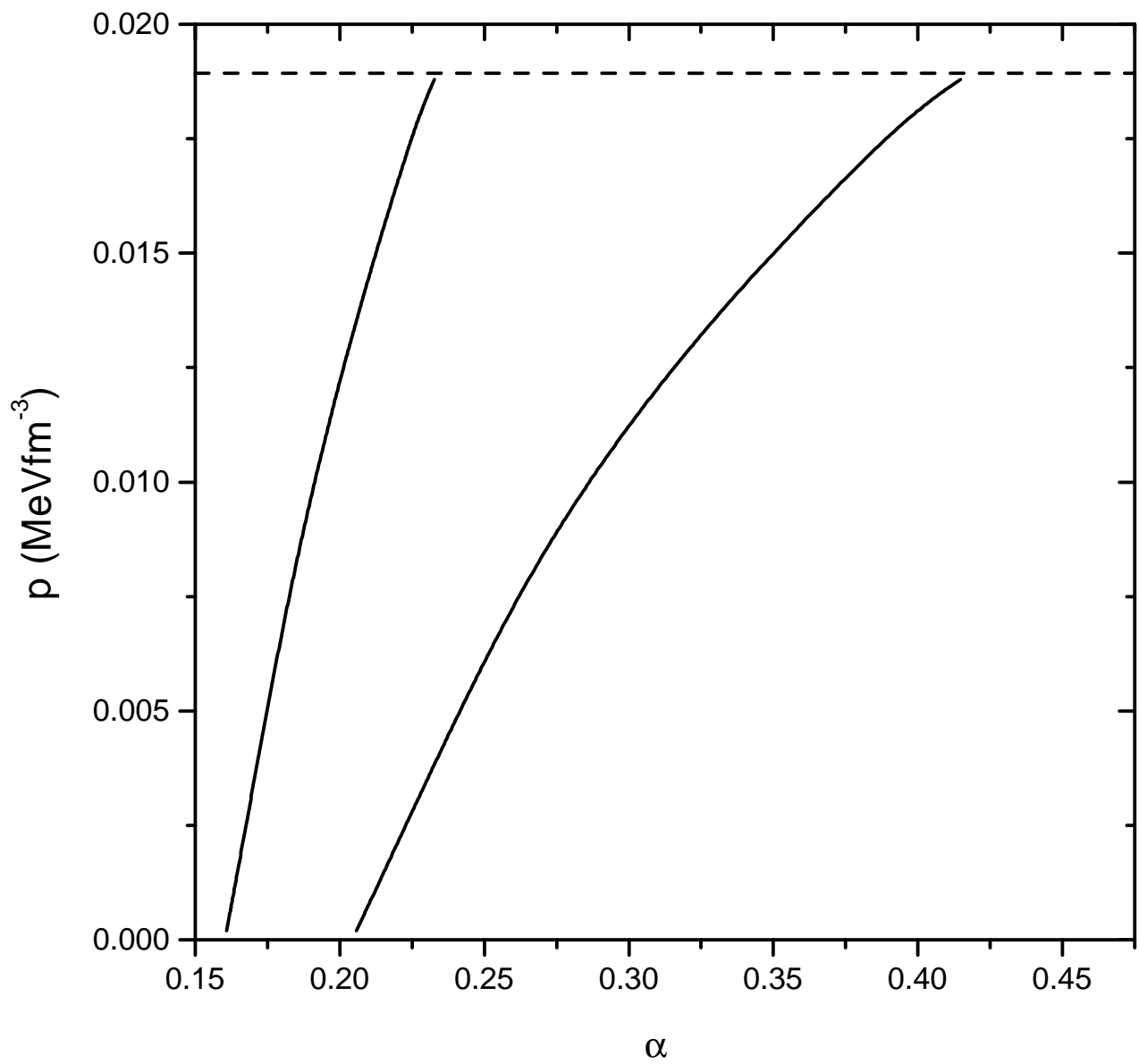
\begin{figure}[h]
\caption{The section of binodal surface for $T=5$ ${\bf MeV}$.}
\label{Fig4}
\end{figure}

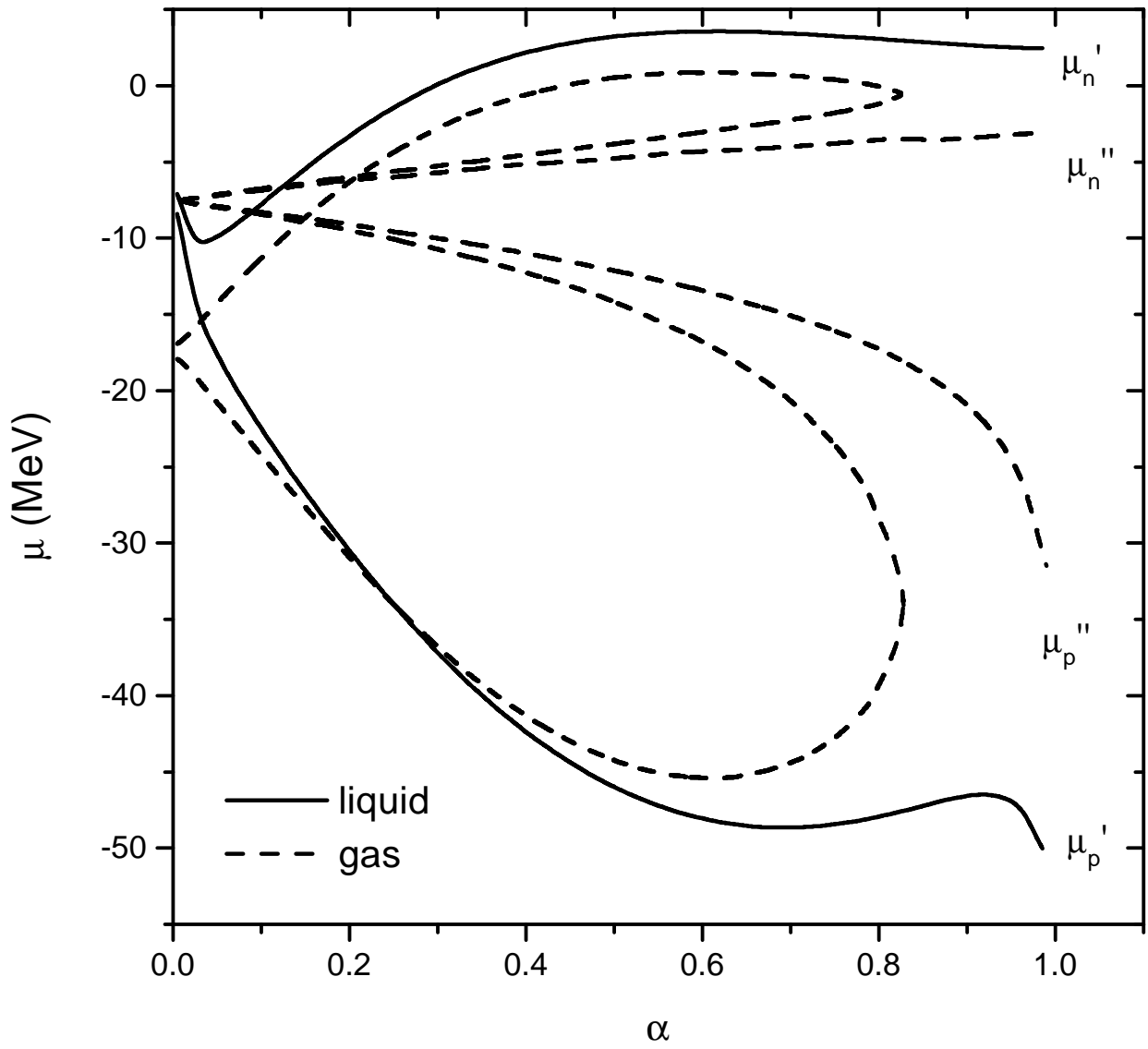
\begin{figure}[h]
\caption{The chemical isobars vs. ${\alpha}$ curves with surface effect only.
}
\label{Fig5}
\end{figure}

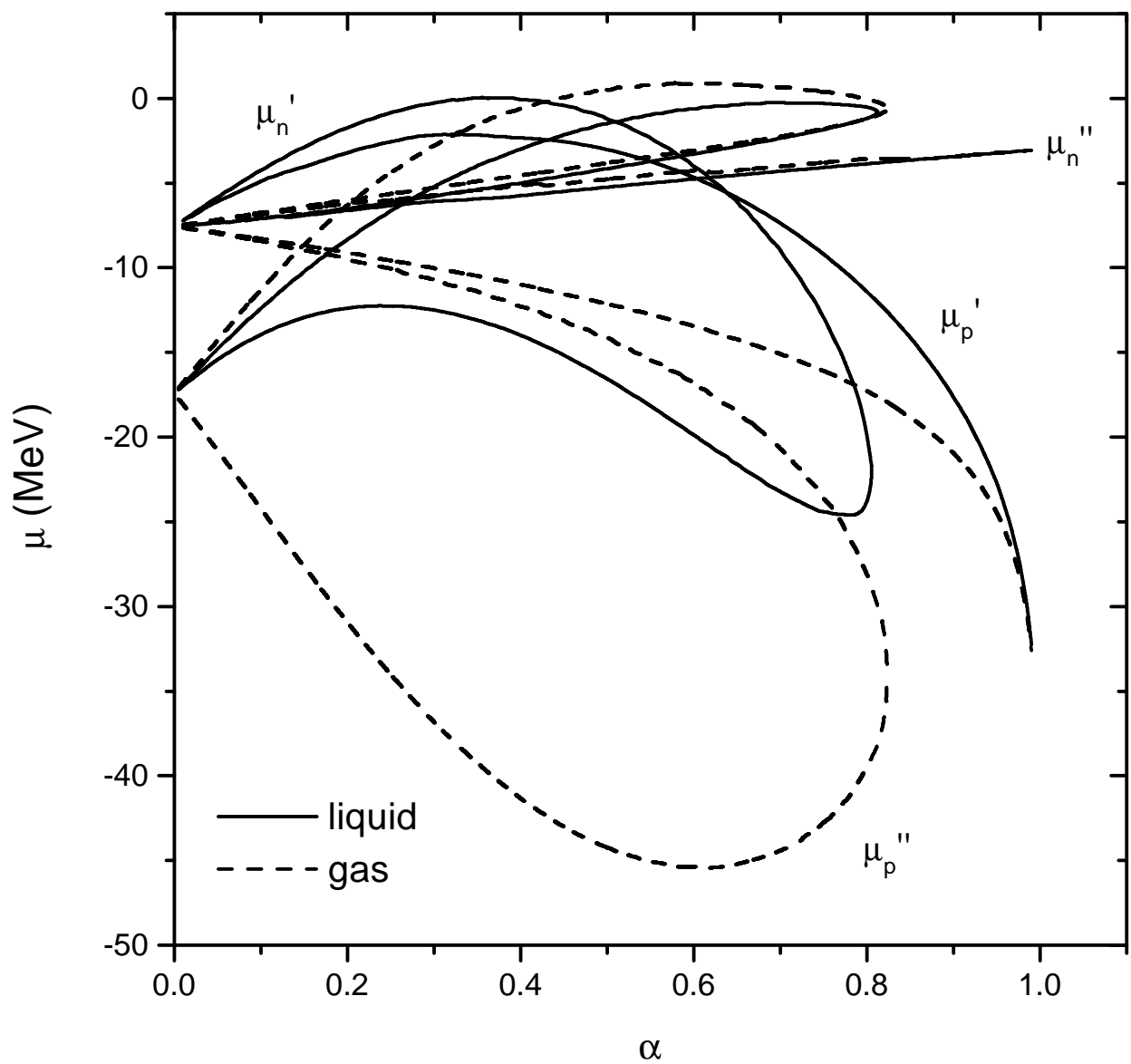
\begin{figure}[h]
\caption{The chemical isobars vs. ${\alpha}$ curves with Coulomb interaction
only.}
\label{Fig6}
\end{figure}

\end{document}